\documentclass[prl,twocolumn,aps,superscriptaddress,longbibliography]{revtex4}
\usepackage{graphicx}
\usepackage[fleqn]{amsmath}
\usepackage{amssymb}
\usepackage[dvipsnames]{xcolor}
\usepackage[normalem]{ulem}
\definecolor{goodgreen}{rgb}{0.1,0.5,0}
\definecolor{goodred}{rgb}{0.7,0,0}
\usepackage{float}
\usepackage{multirow}
\usepackage[colorlinks,urlcolor=goodgreen,citecolor=blue,linkcolor=goodred]{hyperref}
\usepackage{bm}
\allowdisplaybreaks
\begin{document}

\title{Nanomechanical vibrational response from electrical mixing measurements}
\author{C. Samanta}
\affiliation{ICFO - Institut De Ciencies Fotoniques, The Barcelona Institute of Science and Technology, 08860 Castelldefels (Barcelona), Spain}
\author{D. A. Czaplewski}
\affiliation{Center for Nanoscale Materials, Argonne National Laboratory, Argonne, IL, 60439, USA}
\author{S. L. De Bonis}
\affiliation{ICFO - Institut De Ciencies Fotoniques, The Barcelona Institute of Science and Technology, 08860 Castelldefels (Barcelona), Spain}
\author{C. B. M{\o}ller}
\affiliation{ICFO - Institut De Ciencies Fotoniques, The Barcelona Institute of Science and Technology, 08860 Castelldefels (Barcelona), Spain}
\author{R. Tormo Queralt}
\affiliation{ICFO - Institut De Ciencies Fotoniques, The Barcelona Institute of Science and Technology, 08860 Castelldefels (Barcelona), Spain}
\author{C. S. Miller}
\affiliation{Center for Nanoscale Materials, Argonne National Laboratory, Argonne, IL, 60439, USA}
\author{Y. Jin}
\affiliation{C2N, CNRS, Université Paris-Saclay, Palaiseau, France}
\author{F. Pistolesi}
\affiliation{Universite de Bordeaux, CNRS, LOMA, UMR 5798, F-33400 Talence, France}
\author{A. Bachtold}
\affiliation{ICFO - Institut De Ciencies Fotoniques, The Barcelona Institute of Science and Technology, 08860 Castelldefels (Barcelona), Spain}

\begin{abstract}

Driven nanomechanical resonators based on low-dimensional materials are routinely and efficiently detected with electrical mixing measurements. However, the measured signal is a non-trivial combination of the mechanical eigenmode displacement and an electrical contribution, which makes the extraction of the driven mechanical response challenging. Here, we report a simple yet reliable method to extract solely the driven mechanical vibrations by eliminating the contribution of pure electrical origin. This enables us to measure the spectral mechanical response as well as the driven quadratures of motion. We further show how to calibrate the measured signal into units of displacement. Additionally, we utilize the pure electrical contribution to directly determine the effective mass of the measured mechanical mode. Our method marks a key step forward in the study of nanoelectromechanical resonators based on low-dimensional materials in both the linear and the nonlinear regime.

\end{abstract}

\maketitle

\vspace{4.0 cm}

\section{I Introduction}
\label{sec:introduction}

Nanomechanical resonators~\cite{Bachtold2022} are exquisite sensors of mass adsorption~\cite{Yang2006,Chaste2012,Malvar2016} and external forces~\cite{Gavartin2012,Bonis2018,Heritier2018,Sahafi2020}. These sensing capabilities  enable advances in different research fields, such as mass spectrometry~\cite{Hanay2012}, surface science~\cite{Wang2010,Yang2011,Noury2019}, heat transport~\cite{Morell2019,Dolleman2020}, in-situ nanofabrication~\cite{Gruber2019}, magnetic resonance imaging~\cite{Degen2009,Rose2018,Grob2019}, scanning probe microscopy~\cite{Li2007,Lepinay2016,Rossi2016}, nanomagnetism~\cite{Losby2015,Rossi2019,Siskins2020,Jiang2020}, and probing viscosity in liquids~\cite{Gil-Santos2015}. 
Many of these studies are carried out with mechanical resonators based on low-dimensional materials, such as carbon nanotubes~\cite{Reulet2000,sazonova2004tunable}, because of their tiny mass. 
However, the detection of motion becomes increasingly difficult as resonators get smaller.

The electrical detection of resonators based on low-dimensional materials is usually realized with a mixing-based method ~\cite{Knobel2003,sazonova2004tunable}, where the vibrations are driven near resonance frequency and detected at a low frequency within the $RC$ bandwidth of the circuit. This down-conversion of the frequency is crucial, since the resonance frequency of the vibrations is usually much larger than the bandwidth imposed by the resistance of the sample and the capacitance of the electrical cables that connect the device to the measurement instruments. Another reason for this frequency down-conversion is to filter out the parasitic background signal of the drive that overwhelms the measured signal of the vibrations; the direct capacitive signal transduction without this mixing rarely works for nanoresonators in contrast to micro- and macro-scale resonators.  

The electrical mixing detection has been applied to resonators based on carbon nanotubes~\cite{sazonova2004tunable,witkamp2006bending,chiu2008atomic,Wang2010,lassagne2009coupling,gouttenoire2010,wu2011capacitive,eichler2011nonlinear,eichler2011parametric,laird2012high,eichler2013symmetry,moser2013,lee2013carbon,moser2014nanotube,benyamini2014real,schneider2014,Bonis2018,kumar2018mechanical,Noury2019,Khivrich2019,rechnitz2021mode}, graphene~\cite{chen2009performance,zande2010large,eichler2011nonlinear,singh2010probing,singh2012coupling,miao2014graphene,parmar2015dynamic,verbiest2018detecting,Luo2018,jung2019ghz,zhang2020coherent,verbiest2021tunable}, transition metal dichalcogenides (TMDs)~\cite{samanta2015nonlinear,yang2016all,yang2017local,samanta2018tuning,manzeli2019self,sengupta2010electromechanical}, and semiconducting nanowires~\cite{solanki2010tuning,mile2010plane,bargatin2005sensitive,bargatin2007efficient,he2008self,fung2009radio,koumela2013high,Sansa2012}.  Different variants of the mixing method were developed by applying either two signals~\cite{sazonova2004tunable} on the device or one signal that is amplitude~\cite{zande2010large} or frequency~\cite{gouttenoire2010} modulated. The transduction from displacement into current can be based on capacitive~\cite{sazonova2004tunable} or piezo-resistive measurements~\cite{Sansa2014}.  Methods were also implemented to measure thermal vibrations~\cite{moser2013} and ring-downs~\cite{schneider2014,Urgell2020} at temperatures down to below $0.1$~K. The fundamental detection limit was theoretically investigated in Ref.~\cite{Wang2017}. Despite this large amount of work, the measurement of the spectral response of nanomechanical vibrations to a driving force -- the most common method to study mechanical resonators~\cite{Bachtold2022} -- remains to be demonstrated with the mixing detection.

Here, we report on a simple, yet reliable, method to measure the spectral mechanical response to a driving force using the mixing method with two signals applied to the device. By properly tuning the phase of the measured signal, we are able to separate the signal of the mechanical vibrations from the signal of pure electrical origin inherent to the mixing method. Moreover, we use the pure electrical contribution as a resource to measure the mass of the mechanical eigenmode. The mass is a key parameter of mechanical resonators, but its determination is challenging, especially for resonators based on low-dimensional materials.

\section{II Device and experimental approach}
\label{sec:device}

\begin{figure}
    \centering
    \includegraphics[scale=1]{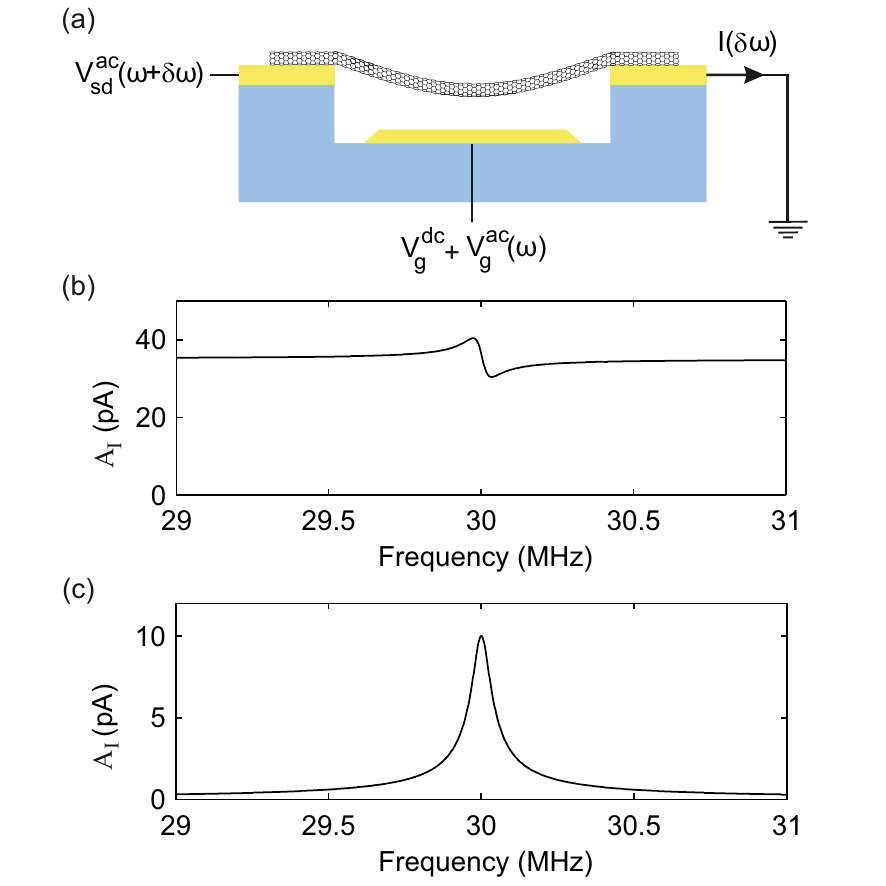}
    \caption{(a) Schematics of measured device. The nanotube is suspended over a gate electrode and electrically connected to two metal electrodes. Two oscillating voltage signals are applied to the device. The current $I$ is measured with a $RLC$ resonator and a low-temperature amplifier~\cite{Bonis2018}. (b) 
    Calculated response of a underdamped harmonic oscillator electromechanically driven by a capacitive force $F(\omega)\propto V_\mathrm{g}^\mathrm{ac}(\omega)$ expected from the mixing method in the limit where the mechanical displacement is much smaller than $C_\mathrm{g} V_\mathrm{g}^\mathrm{ac}/C_\mathrm{g}'V_\mathrm{g}^\mathrm{dc}$. (c) Same as panel b but in the opposite limit.
    }
    \label{fig_1}
\end{figure}

We produce nanotube mechanical resonators by growing nanotubes using chemical vapor deposition on prepatterned electrodes. The nanotube is suspended $\simeq 150$~nm above a gate electrode and connected between two metal electrodes~\cite{Bonis2018} (Fig.~\ref{fig_1}a). We clean the nanotube surface from contamination molecules by applying a large current through the device under vacuum at low temperature~\cite{Yang2020}. 

We detect the vibrations of the nanotube resonator by capacitively driving it with an oscillating voltage $V_\mathrm{g}^\mathrm{ac}\cos\omega t$ on the gate electrode, applying the voltage $V_\mathrm{s}^\mathrm{ac}\cos{((\omega+\delta\omega)t+\varphi_\mathrm{e})}$ on the source electrode, and measuring the current at frequency $\delta\omega$ from the drain electrode with a lock-in amplifier \cite{Bonis2018} (Fig.~\ref{fig_1}a) where $\varphi_\mathrm{e}$ is the phase difference between the two oscillating voltages. We set $\delta\omega$ within the bandwidth of the circuit and we sweep $\omega$ through the mechanical frequency $\omega_\mathrm{m}$ ($\delta\omega\ll\omega_\mathrm{m}$). All the measurements are carried out with the device in the single-electron tunneling regime~\cite{Armour2004,Clerk2005,Pistolesi2007,Usmani2007,Micchi2015} at the temperature $T=6$~K.

To detect the vibrations, the nanotube has to behave as a transistor such that the conductance $G$ depends on the charge $Q$ in the nanotube. The application of $V_\mathrm{g}^\mathrm{ac}\cos\omega t$ modulates the charge through two terms $\delta Q=C_\mathrm{g} \delta V_\mathrm{g}+\delta C_\mathrm{g} V_\mathrm{g}$. The first term has a pure electrical origin, while the second term is proportional to the driven vibration displacement $\delta z$ via $\delta C_\mathrm{g}=C^{\prime}_\mathrm{g}\delta z$, where $C^{\prime}_\mathrm{g}$ is the spatial derivative of the capacitance.
The application of $V_\mathrm{s}^\mathrm{ac}\cos{((\omega+\delta\omega)t+\varphi_\mathrm{e})}$ enables one to mix down the modulation of $G$ into a current oscillation at the frequency $\delta\omega$ within the circuit bandwidth via Ohm$'$s law $I=GV_\mathrm{s}$. The mixing intertwines the two terms of the charge modulation. As a result, the displacement of the vibrations driven at frequency $\omega$ and the current at frequency $\delta\omega$ given by    
\begin{align}
\label{eq:displacementz}
&z= A_\mathrm{z} \cos(\omega t+\phi_\mathrm{z})= X_\mathrm{z} \cos\omega t+ Y_\mathrm{z} \sin\omega t,\\
\label{eq:displacementI}
&I= A_\mathrm{I} \cos(\delta\omega t+\phi_\mathrm{I})=X_\mathrm{I} \cos\delta\omega t+ Y_\mathrm{I} \sin\delta\omega t.
\end{align}
are related in a cumbersome way, since the quadratures $X_\mathrm{I}$ and $Y_\mathrm{I}$ of the current depend on the quadratures $X_\mathrm{z}$ and $Y_\mathrm{z}$ of the displacement as
\begin{align}
\label{eq:conversionX}
&
X_\mathrm{I} = \alpha \left[(X_\mathrm{z} + 
C_\mathrm{g} V_\mathrm{g}^\mathrm{ac}/C_\mathrm{g}'V_\mathrm{g}^\mathrm{dc})
\cos{\varphi_\mathrm{e}}
-Y_\mathrm{z}\sin{\varphi_\mathrm{e}}\right],
\\
\label{eq:conversionY}
&Y_\mathrm{I}= \alpha \left[-(X_\mathrm{z} + 
C_\mathrm{g} V_\mathrm{g}^\mathrm{ac}/C_\mathrm{g}'V_\mathrm{g}^\mathrm{dc})
\sin{\varphi_\mathrm{e}}+Y_\mathrm{z}\cos{\varphi_\mathrm{e}}\right],
\end{align}
with 
$\alpha= 
({\partial G}/{\partial V_\mathrm{g}}) V_\mathrm{s}^\mathrm{ac} V_\mathrm{g}^\mathrm{dc}C'_\mathrm{g} /2 C_\mathrm{g}$,
$({\partial G}/{\partial V_\mathrm{g}})$ the transconductance, and $V_\mathrm{g}^\mathrm{dc}$ the static voltage applied to the gate (Appendix A). We note that the work function difference between the nanotube and the gate electrode has to be subtracted from $V_\mathrm{g}^\mathrm{dc}$. 
%
%

The downside of the mixing method is that the measured current is not directly proportional to the driven vibration displacement. 
The amplitude of the current is given by 
$A_\mathrm{I} =  \alpha\sqrt{( X_\mathrm{z} + C_\mathrm{g} V_\mathrm{g}^\mathrm{ac}/C_\mathrm{g}'V_\mathrm{g}^\mathrm{dc})^2  + Y_\mathrm{z}^2 }$.
In the limit where the displacement $z$ is much smaller than 
$ C_\mathrm{g} V_\mathrm{g}^\mathrm{ac}/C_\mathrm{g}'V_\mathrm{g}^\mathrm{dc}$,
the response of $A_\mathrm{I}$ consists of a signal proportional to $X_\mathrm{z}$ together with a large, frequency-independent background that has a pure electrical origin, see Fig.~\ref{fig_1}b. In the opposite limit, the responses of $A_\mathrm{I}$ and $A_\mathrm{z}$ become proportional to each other (Fig.~\ref{fig_1}c).

\begin{figure}
    \centering
    \includegraphics[scale=1]{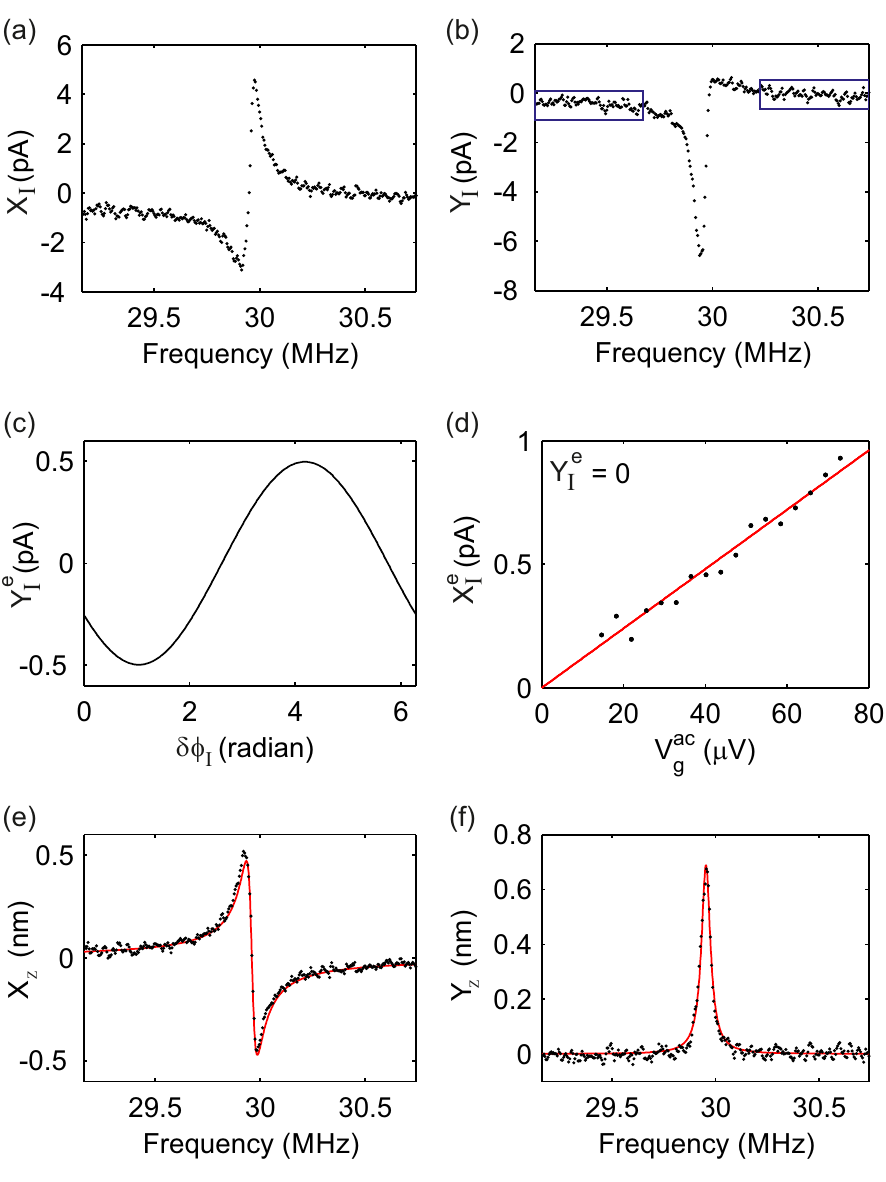}
    \caption{(a,b) Spectral response of the current quadratures $X_\mathrm{I}$ and $Y_\mathrm{I}$ to the driven capacitive force. The two blue boxes indicate the $Y_\mathrm{I}$ values used to compute the background offset $Y_\mathrm{I}^\mathrm{e}$. (c) Estimated background offset $Y_\mathrm{I}^\mathrm{e}$ from the data in a and b by incrementing the phase $\phi_\mathrm{I}$ in Eq.~\ref{eq:displacementz} by $\delta\phi_\mathrm{I}$. (d) Background current $X_\mathrm{I}^\mathrm{e}$ with pure electrical origin with $\delta\phi_\mathrm{I}$ set so that $Y_\mathrm{I}^\mathrm{e}=0$. The red line is a linear fit of the data. (e,f) Spectral response of the displacement quadratures $X_\mathrm{z}$ and $Y_\mathrm{z}$ to the driven capacitive force after having subtracted $X_\mathrm{I}^\mathrm{e}$ from $X_\mathrm{I}$. The data are compared to the quadratures expected for a linear oscillator (red lines).     }
    \label{fig_2}
\end{figure}

\begin{figure} [t]
    \includegraphics[scale=1]{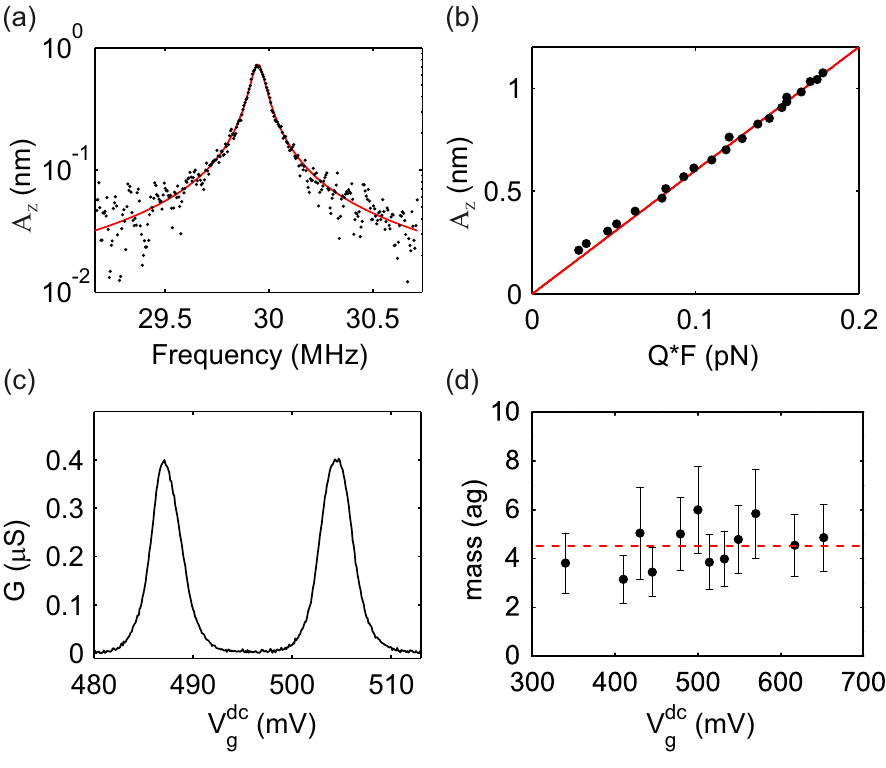}
    \caption{(a) Spectral response of the displacement amplitude $A_\mathrm{z}$ to the driven capacitive force after having subtracted $X_\mathrm{I}^\mathrm{e}$ from $X_\mathrm{I}$. The data are compared to a Lorentzian peak (red line). (b) Force response of the displacement amplitude $A_\mathrm{z}$ at the mechanical resonance frequency. The red line is a linear fit of the data. The force is multiplied by the quality factor, since the latter varies when increasing the driving force~\cite{eichler2011nonlinear}. (c) Electrical conductance of the nanotube device as a function of gate voltage. (d) Mass of the eigenmode measured at different gate voltage values. The red dashed line indicates the average mass of $4.5$~ag.}
    \label{fig_3}
\end{figure}

It is possible to separate the current signal of pure electrical origin by setting $X_\mathrm{I}\propto X_\mathrm{z} + C_\mathrm{g} V_\mathrm{g}^\mathrm{ac}/C_\mathrm{g}'V_\mathrm{g}^\mathrm{dc}$ and $Y_\mathrm{I}\propto Y_\mathrm{z}$  by properly adjusting the phase $\phi_\mathrm{LIA}$ of the lock-in amplifier, which enters  Eqs.~\ref{eq:conversionX},\ref{eq:conversionY} by replacing $\varphi_\mathrm{e}\rightarrow\varphi_\mathrm{e}-\phi_\mathrm{LIA}$.
However, this is not practical, since the phase of the lock-in amplifier  often needs to be readjusted when changing $V_\mathrm{g}^\mathrm{ac}$ and $V_\mathrm{g}^\mathrm{dc}$.
Alternatively, the signal of pure electrical origin can be separated after the measurements by performing a rotation of the 
angle $\phi_\mathrm{I}$ in the plane $(X_\mathrm{I},Y_\mathrm{I})$
for the data. 
This is equivalent to the transformation $\varphi_\mathrm{e}\rightarrow\varphi_\mathrm{e}-\phi_\mathrm{LIA}$. 
%
To illustrate this alternative method, we proceed with the response of the two quadratures $X_\mathrm{I}$ and $Y_\mathrm{I}$ of the current directly acquired from the lock-in amplifier (Figs.~\ref{fig_2}a,b). 
The two responses cannot be described by the usual functional forms of driven linear oscillators, since the phase of the lock-in amplifier was not adjusted beforehand. 
We then compute the background offset of $Y_\mathrm{I}$ by incrementing the rotation phase $\phi_\mathrm{I}$ by $\delta\phi_\mathrm{I}$ from $0$ to $2\pi$ (Fig.~\ref{fig_2}c). 
When this background offset in $Y_\mathrm{I}$ is zero, all the current signal of pure electrical origin is in $X_\mathrm{I}$ and can be subtracted from the data. 
The resulting quadrature responses have now the familiar functional form of linear oscillators (Figs.~\ref{fig_2}e,f) and the spectral response of the displacement is well described by a Lorentzian (Fig.~\ref{fig_3}a).

\section{III Results and discussion}
\label{sec:results}

We use the subtracted background current $X_\mathrm{I}^\mathrm{e}$ of pure electrical origin to calibrate the displacement of the nanotube resonator in units of meters (Fig.~\ref{fig_3}a). This background current is given by 
$X_\mathrm{I}^\mathrm{e}=({\partial G}/{\partial V_\mathrm{g}}) V_\mathrm{s}^\mathrm{ac} V_\mathrm{g}^\mathrm{ac}/2$;
we verify that it  depends linearly on $V_\mathrm{g}^\mathrm{ac}$ (Fig.~\ref{fig_2}d). 
The two quadratures then read:
\begin{align}
\label{eq:computeddispacementquadratures}
&
X_\mathrm{z}= 
\frac{C_\mathrm{g}V_\mathrm{g}^\mathrm{ac} }{ C'_\mathrm{g}  V_\mathrm{g}^\mathrm{dc} }
\frac{X_\mathrm{I}-X_\mathrm{I}^\mathrm{e}}
{X_\mathrm{I}^\mathrm{e}},
\quad
Y_\mathrm{z}= 
\frac{C_\mathrm{g}V_\mathrm{g}^\mathrm{ac} }{ C'_\mathrm{g}  V_\mathrm{g}^\mathrm{dc} }
\frac{Y_\mathrm{I}}{X_\mathrm{I}^\mathrm{e}}.
\end{align}
The calibration of the displacement is subject to the uncertainty in the estimation of $C_\mathrm{g}/C'_\mathrm{g}$ (see below).

The current $X_\mathrm{I}^\mathrm{e}$ of pure electrical origin also enables quantifying the mass of the mechanical mode in a way that is simple and reliable. In Fig.~\ref{fig_3}b, we compute the force response of the displacement amplitude at resonance frequency $\omega_\mathrm{m}$ in the linear regime using 
\begin{align}
\label{eq:displacementamplitudeforce}
&A_\mathrm{z}= \frac{C_\mathrm{g}}{C^{\prime}_\mathrm{g}}\frac{V_\mathrm{g}^\mathrm{ac}}{V_\mathrm{g}^\mathrm{dc}}\frac{Y_\mathrm{I}}{X_\mathrm{I}^\mathrm{e}},
\quad
F=\beta C^{\prime}_\mathrm{g}V_\mathrm{g}^\mathrm{dc}V_\mathrm{g}^\mathrm{ac},
\end{align}
where $Y_\mathrm{I}$ corresponds to the current amplitude at resonance frequency after having separated the signal of pure electrical origin. The constant $\beta$ can be different from one for electron transport in the single-electron regime (Eq.~\ref{betaDef} and Appendix A).
The mass $m$ is determined from the slope of the force-displacement response using $A_\mathrm{z}=(Q/m\omega_\mathrm{m}^2)F$ with $Q$ the quality factor. The slope depends on the current terms $Y_\mathrm{I}$ and $X_\mathrm{I}^\mathrm{e}$  measured from the lock-in amplifier, but is independent of $V_\mathrm{g}^\mathrm{ac}$, $V_\mathrm{s}^\mathrm{ac}$, and ${\partial G}/{\partial V_\mathrm{g}}$ that enter the prefactor $\alpha$ in the current-displacement conversion in Eqs.~\ref{eq:conversionX},\ref{eq:conversionY} and whose values could be somewhat altered by the amplification chain and the losses along the coaxial cables. We determine $m=4.5\pm1.5$~ag from the mass measured at different $V_\mathrm{g}^\mathrm{dc}$ values  (Fig.~\ref{fig_3}d). This value is consistent with the length of the suspended nanotube measured by scanning electron microscopy and assuming a $1.5$~nm radius single-wall nanotube. 

The uncertainty in the mass measurement comes from the uncertainty in the estimation of the nanotube-gate separation $d$ and the mass fluctuations in Fig.~\ref{fig_3}d. The separation $d=150\pm20$~nm measured by atomic force microscopy enters in the estimation of ${C^{\prime}_\mathrm{g}}={C_\mathrm{g}}/{d \ln({{2d}/{r}})}$ in Eqs.~\ref{eq:computeddispacementquadratures},\ref{eq:displacementamplitudeforce} when considering the capacitance between a tube with radius $r$ separated from a plate by the distance $d$. We estimate $C_\mathrm{g}=e/\Delta V_\mathrm{g}=9.7$~aF from the separation $\Delta V_\mathrm{g}$ in gate voltage between two conductance peaks associated with single-electron tunneling (Fig.~\ref{fig_3}c). This capacitance is consistent with $C_\mathrm{g}=12.9$~aF obtained from the device geometry measured by scanning electron microscopy and atomic force microscopy.  
The fluctuations of $m$ in Fig.~\ref{fig_3}d are partly due to the error in the estimation of the average charge occupation $f$, which varies between 0 and 1 when sweeping $V_\mathrm{g}^\mathrm{dc}$ through the conductance peaks (Fig.~\ref{fig_3}c), since $f$ enters in the prefactor $\beta$ of the driven force in Eq.~\ref{eq:displacementamplitudeforce} as
\begin{equation}
\beta=1-\dfrac{C_\mathrm{g}}{C_\mathrm{\Sigma}} +f(1-f)\dfrac{C_\mathrm{g}}{C_\mathrm{\Sigma}} \dfrac{e^2/C_\mathrm{\Sigma}}{k_\mathrm{B}T}, \label{betaDef}
\end{equation}
in the incoherent single-electron tunneling regime. Here, $C_{\Sigma}$ is the total capacitance of the single-electron transistor and varies gradually from $19.9$~aF to $26.5$~aF when sweeping $V_\mathrm{g}^\mathrm{dc}$ over multiple conductance peaks. The fluctuations of $m$ are also attributed to the slow increase of the contamination on the nanotube surface at 6~K; the three largest $m$ values in Fig.~\ref{fig_3}d are obtained from force-displacement measurements carried out one month after the first measurements.      

\section{IV Conclusion}
\label{sec:conclusion}

In summary, we show how to measure the spectral mechanical response using electrical mixing measurements. Our method enables us to calibrate the displacement in meters. Another asset of this method is the determination of the mass of the measured mechanical eigenmode, which is a key parameter of the mechanical resonators when used in sensor applications. This work opens the possibility to quantitatively study nanoelectromechanical resonators in the nonlinear regime, where different mesoscopic phenomena can be explored~\cite{Bachtold2022}. In previous studies, the shape of the nonlinear response measured with the mixing method was often complicated and it was not possible to unambiguously separate the contribution of the nonlinear mechanical response from the contribution with a pure electrical origin. By contrast, it is expected that our method is able to extract the nonlinear mechanical response in a straightforward way.  

\section{V Acknowledgements}
\label{sec:Acknowledgements}
We acknowledges ERC Advanced Grant No. 692876 and MICINN Grant No. RTI2018-097953-B-I00. Work performed at the Center for Nanoscale Materials, a U.S. Department of Energy Office of Science User Facility, was supported by the U.S. DOE, Office of Basic Energy Sciences, under Contract No. DE-AC02-06CH11357. We also acknowldge AGAUR (Grant No. 2017SGR1664), the Fondo Europeo de Desarrollo, the Spanish Ministry of Economy and Competitiveness through Quantum CCAA and CEX2019-000910-S [MCIN/ AEI/10.13039/501100011033], Fundacio Cellex, Fundacio Mir-Puig, Generalitat de Catalunya through CERCA, the French Agence Nationale de la Recherche (Grant No. SINPHOCOM ANR-19-CE47-0012), Marie Sk\l odowska-Curie (Grant No. 101023289).

\section{Appendix A: Two-source mixing method}
\label{sec:two-source}

We consider a double-clamped mechanical resonator that is capacitively coupled to an immobile gate electrode. 
The two-source mixing method requires that the conductance through the resonator varies when sweeping the gate voltage. In what follows, we consider the regime of single-electron tunneling, but the same final result for the mixing current is obtained for any other regime.
The vibrations are driven by applying an oscillating voltage $V_\mathrm{g}^\mathrm{ac}\cos\omega t$ on the gate electrode. 
When applying the voltage $V_\mathrm{s}^\mathrm{ac}\cos{((\omega+\delta\omega)t+\varphi_\mathrm{e})}$ on the source electrode, the mixing current at frequency $\delta\omega$ arises in the Taylor expansion of the current in $z$ and $V_\mathrm{g}$ \cite{sazonova2004tunable}.
The dependence of the current on these two quantities can be traced back to the tunnelling rate dependence  on the electrostatic energy  difference between the two relevant charge states of the dot. 
For vanishing bias voltage the electrostatic energy reads 
$E_E(Q)=(Q+C_g(z) V_g)^2/2C_\Sigma(z)$ with $C_g$ and $C_\Sigma$ the gate and total capacitances
and $Q$ the charge on the dot.
This gives for the relevant energy difference 
$\Delta E = E_E(Q-e)-E_E(Q)=e(e/2-Q-C_\mathrm{g}(z) V_\mathrm{g})/C_\Sigma(z)$.
The current is then a function of $\Delta E(z,V_\mathrm{g})$.
Expanding the current expression 
for $e V_\mathrm{s}\ll k_B T$, small displacement $z$ [given by Eq.~(\ref{eq:displacementz})], and $V_\mathrm{g}^\mathrm{ac}$ one obtains:
\begin{align}
\label{eq:currentterms}
&{I}=\frac {\partial G}{\partial V_\mathrm{g}} V_\mathrm{s}^\mathrm{ac}\cos{((\omega+\delta\omega)t+\varphi_\mathrm{e})}\nonumber\\
&\times[V_\mathrm{g}^\mathrm{ac}\cos\omega t+V_\mathrm{g}^\mathrm{dc} C_\mathrm{g}'/C_\mathrm{g}
[X_\mathrm{z} \cos (\omega t)+Y_\mathrm{z} \sin(\omega t)].
\end{align}
Here ${\partial G}/{\partial V_\mathrm{g}}$ is the transconductance of the nanotube device, 
$V_\mathrm{g}^\mathrm{dc}$ the static voltage applied to the gate and 
we assumed $Q \approx -C_\mathrm{g} V_\mathrm{g} \gg e$.
Expanding the argument of the first cosine and averaging over a period $2\pi/\omega$ gives
the mixing current $I^\mathrm{\delta\omega}$ at frequency $\delta\omega$:
\begin{align}
\label{eq:currentterms}
I^\mathrm{\delta\omega} &=
{1\over 2} \frac {\partial G}{\partial V_\mathrm{g}} V_\mathrm{s}^\mathrm{ac}
\left[ 
\cos{(\delta\omega t +\varphi_\mathrm{e})} 
\left(
V_\mathrm{g}^\mathrm{ac} +\frac{V_\mathrm{g}^\mathrm{dc} C_\mathrm{g}'}{C_\mathrm{g} }X_\mathrm{z}
\right)
\right.
\nonumber \\
& \left.
-
\sin{(\delta\omega t +\varphi_\mathrm{e})} \frac{V_\mathrm{g}^\mathrm{ac} C_\mathrm{g}'}{C_\mathrm{g} }Y_\mathrm{z}
\right].
\end{align}
This leads to the mixing current quadratures $X_\mathrm{I}$ and $Y_\mathrm{I}$ in Eqs.~\ref{eq:conversionX},\ref{eq:conversionY}. The expression of the mixing current in Eq.~\ref{eq:currentterms} is the same for other types of conductors, such as the electronic Farby-P\'{e}rot interferometer or the the field-effect transistor~\cite{sazonova2004tunable}. In the next appendix, we show that the capacitive force in the single-electron tunneling regime is different from that in other regimes.

\section{Appendix B: driving force in the single-electron tunneling regime}
\label{sec:two-source}

\newcommand{\beq}{\begin{equation}}
\newcommand{\eeq}{\end{equation}}

We discuss here the oscillating force acting on a mechanical resonator hosting a dot that behaves as a single-electron transistor in the limit typically 
realized in experiments with a slow oscillator $\Gamma \gg \omega_m$, where
$\Gamma$ is the typical incoherent tunneling rate ($k_B T \gg \hbar \Gamma$).
When the gate voltage is modulated, the charge on the dot changes, leading to 
an additional oscillating force acting on the oscillator. This is the reason why the constant $\beta$ in Eq.~\ref{eq:displacementamplitudeforce} for the capacitive force can deviate from one.
The total capacitive force between the resonator and the the gate electrode can be written as 
\beq
	F=
	-{\partial \over \partial z} {Q_g^2 \over 2 C_g(z)}  	
	= {Q_g^2  C'_g\over 2 C_g^2 }  
	\label{eq:Fg}
\eeq
where $Q_g$ is the charge on the gate electrode (we assume that the capacitances to 
the source or drain are not modified by the displacement of the resonator). 
In the sequential tunnelling regime the charge on the dot is always an integer  multiple 
of the elementary charge $Q=-e(n_0+ n)$, with 
$n_0$ and $n$ integers, and only $n$ varies between 0 and 1.
From electrostatics the gate charge is then:
\beq
	Q_g= C_g V_g - {C_g \over C_\Sigma} (V_s C_s + C_d V_d+ C_g V_g + Q).
\eeq
where we introduced the source and drain voltages ($V_s$,$V_d$) and capacitances ($C_s$, $C_d$) 
with $C_\Sigma=C_g+C_s+C_d$. 
Since the number of electrons fluctuates of one unit during transport, 
there are actually two forces acting on the dot, one for each value of $Q$. 
Using the separation of time scales we can assume that the oscillator cannot respond to the fast electron fluctuations, and 
thus it feels an average force given by the average value of $Q$.
When $\delta V_g(t)=V_g^\mathrm{ac}\cos(\omega t)$ is applied to the gate electrode, we can write that the resulting variation of the charge on the gate electrode reads:
\beq
	\delta Q_g= C_g \delta V_\mathrm{g} 
	\left[ 1 - {C_g \over C_\Sigma}  \right]  +   {C_g \over C_\Sigma}  \delta \langle Q\rangle
		\label{eq:deltaQg}
	.
\eeq
We can neglect the higher orders in the $z$-dependence of the capacitance when computing the force in Eq.~\ref{eq:Fg}, since this gives rise only to a renormalization of the resonance frequency.
The variation of $Q$ is controlled by the master equation for the charge. 
Assuming that only two charge states are 
possible, one has
$
	\langle Q \rangle = -n_0 e -e f
$
with $f$ the Fermi function
$
	f=(e^{{\varepsilon/k_B T}}+1)^{-1}
$
where the $\varepsilon$ dependence on the gate voltage 
is 
$\delta \varepsilon = -e C_g \delta V_g/C_\Sigma$. 
We obtain then
\beq
	\delta \langle Q \rangle = -{e^2\over k_B T}   
	{C_g\over C_\Sigma} \delta V_g 
	f(1-f).
	\label{eq:deltaQ}
\eeq
Note that the factor ${(e^2/C_\Sigma) /k_B T}\gg 1$ 
in the Coulomb blockade regime. 
This term $\delta \langle Q \rangle$ is largest for gate voltages at which the peak conductance is highest and where
$f=1/2$.
Inserting Eq.~\ref{eq:deltaQ} into Eq.~\ref{eq:deltaQg} and 
Eq.~\ref{eq:Fg} one obtains Eq.~\ref{betaDef} of the main text. 

For completeness, it can be useful to recall the derivation of the coupling constant
between the mechanical and electronic degrees of freedom.
This is the variation of the force acting on the oscillator when an electron on the dot is added or removed. 
\beq    
    F_0 = F_g(Q)-F_g(Q-e) = 
    { C_g' e^2 \over C_g C_\Sigma}(Q_g/e-1)
    .
\eeq
For $|Q_g|\gg e$ and $V_s\approx V_d \approx 0$ one finds 
\beq    
    F_0 =  
    { C_g' V_g e \over C_\Sigma} 
    .
\eeq


 \vspace{10.0 cm}

\end{document}